\documentclass[aps,prd,twocolumn,showpacs,groupedaddress]{revtex4}  
\usepackage{graphicx}
\usepackage{booktabs}
\usepackage{amssymb,bm,mathrsfs,bbm,amscd}
\usepackage[tbtags]{amsmath}
\usepackage{epstopdf}
\usepackage{color,soul}
\usepackage{siunitx}

\begin{document}
\title{Limits on Lorentz violation in gravity from worldwide superconducting gravimeters}
\author{Cheng-Gang Shao}
\author{Ya-Fen Chen}
\author{Rong Sun}
\author{Lu-Shuai Cao}
\author{Min-Kang Zhou}
\author{Zhong-Kun Hu}\email[E-mail:]{zkhu@mail.hust.edu.cn}

\affiliation{MOE Key Laboratory of Fundamental Physical Quantities
Measurements, Hubei Key Laboratory of Gravitation and Quantum Physics, School of Physics, Huazhong University of Science and Technology, Wuhan 430074, People's Republic of China}

\author{Chenghui Yu}
\author{Holger M\"uller}\email[E-mail:]{hm@berkeley.edu}
\affiliation{Department of Physics, University of California, Berkeley, California 94720, USA}

\date{\today}

\begin{abstract}
We have investigated Lorentz violation through analyzing tides-subtracted gravity data measured by superconducting gravimeters. At the level of precision of superconducting gravimeters, we have brought up and resolved an existing issue of accuracy due to unaccounted local tidal effects in previous solid-earth tidal model used. Specifically, we have taken local tides into account with a brand new first-principles tidal model with ocean tides included, as well as removed potential bias from local tides by using a worldwide array of 12 superconducting gravimeters. Compared with previous test with local gravimeters, a more accurate and competitive bound on space-space components of gravitational Lorentz violation has been achieved up to the order of $10^{-10}$.
\end{abstract}


\maketitle

Einstein's equivalence principle is the foundation of general relativity. It is based on the universality of free fall, local Lorentz invariance and, local position invariance \cite{0, 1}. The universality of free fall has been tested up to accuracies of $10^{-13}$ \cite{1a, 1b, 1c, 1d}. Local position invariance has been tested, e.g., by gravitational red shift measurement with atom interferometers or clocks \cite{2,2a}. In comparison, testing local Lorentz invariance (LLI) is a broad field, as violations of LLI might manifest themselves in the gravity sector itself, or in the matter sectors as well as their coupling \cite{3a,3b}.

In the simplest case, violations of LLI in the gravity sector manifest themselves as a dependence of the force of gravity between two objects on the direction of their separation. Competitive bounds in this sector have been established by various experiments and observations\cite{8a, 8b}, such as gravimetry \cite{4a, 4b, 4c}, lunar laser ranging \cite{5, 5a, 5b} and astrophysics observations \cite{6, 6a}. Among these, local gravimetry is the one of the easy-to-access and very precise ground-based method. The underlying idea is simple: if the force of gravity is anisotropic, then the local acceleration of free fall on the rotating earth should exhibit a modulation correlated with the earth's rotation. In analyzing such tests, the influence of the sun, the moon and the planets have to be taken out, which is done by subtracting a "tidal model" describing of these influences.

However, a persisting problem has been pointed out in previous works \cite{4a, 4b, 4c} whether a simple first-principles solid-earth tidal model or a more sophisticated empirical model should be used. The simple first-principles solid-earth tidal model does not include any Lorentz violation signal. But it's not accurate enough beyond $10^{-10}$g without including local tidal effects like ocean tides. At the precision of superconducting gravimeters, it may produce fake Lorentz violating signals \cite{4c}. Sophisticated empirical models are a lot more accurate, but it's based on fitting of gravity measurement which itself may contain Lorentz-violating signals. In this work, we have reconciled the conflict with a worldwide network of gravimeters analyzed with first-principle tidal models with ocean tides included.   

A worldwide network of local gravity observations was made public through the Global Geodynamics Projects (GGP) which is now called International Geodynamics and Earth Tide Service (IGETS) \cite{18}. We have analyzed gravity data from twelve stations over a period spanning up to 20 years and spanning a total of 120 station-years with a first-principles tidal model which includes ocean tides, yielding accurate and competitive bounds on several modes of Lorentz violation. The large amount of data from multiple stations included in our analysis also enables us to remove potential bias from local tidal effects of specific location. 



We express our limits on Lorentz violation in terms of coefficients in the minimal standard model extension (SME) of the pure-gravity sector. The SME is an effective field theory that offers a theoretical framework founded on well-established physics of the standard model and general relativity to describe experimental observable Lorentz violation (LV) signals \cite{7a, 7b, 7c}. It can be formulated as a Lagrangian density containing general relativity and the minimally coupled standard model, and terms introducing Lorentz violation from different sectors \cite{7d,7e}. The Lorentz-violating terms can be categorized based on its mass dimension $d$. The minimal SME of the pure-gravity sector that we use in this work is a subset of the SME where only the leading-order Lorentz-violating term of mass dimension $d=4$ caused by gravitational fields is taken into account.

The action of minimal SME in the pure-gravity sector can be written as \cite{7d}
\begin{eqnarray}\label{equation1:eps}
S =S_{EH}+S_{LV}+S^{'},  \label{Eq:1}
\end{eqnarray}
where $S_{EH}$ is the Einstein-Hilbert action of general relativity, $S_{LV}$ the leading Lorentz-violating gravitational coupling, $S'$ the general matter action of the standard model. The first two terms can be further formulated as \cite{7d}
\begin{eqnarray}\label{equation2:eps}
S_{EH}+S_{LV} &=&\frac{1}{{16\pi G_N}}\int d^{4}xe[(R-2\Lambda)\nonumber \\
& &+(-uR+{s^{\mu \nu }}R_{\mu \nu }^T+ {t^{\kappa \lambda \mu \nu }}{C_{\kappa \lambda \mu \nu }})],  \label{Eq:2}
\end{eqnarray}
where $u$, ${s^{\mu \nu }}$, ${t^{\kappa \lambda \mu \nu }}$ are the fields contributing to Lorentz violation, $R_{\mu \nu }^T$ is the traceless Ricci tensor, $\Lambda$ the cosmological constant, and $C_{\kappa \lambda \mu \nu }^{}$ the Weyl tensor. In this formulation, the vacuum expectation values of the Lorentz-violating fields $\bar u$, $\bar s^{\mu \nu}$, $\bar t^{\kappa \lambda \mu \nu}$ are signals we are looking for. However, within the post-Newtonian treatment, the coefficients $\bar u$ and $\bar t^{\kappa \lambda \mu \nu }$ could be neglected since $\bar u$ acts as an unobservable rescaling of ${G_N}$ while $\bar t^{\kappa \lambda \mu \nu }$ has no effects on physical experiments at the leading order \cite{7e}. With the only remaining observable fields ${\bar s^{\mu \nu }}$, the Lagrangian of a two-body system can be written as:

\begin{eqnarray}\label{equation3:eps}
L^T &=& \frac{1}{2}m{v^2} + G\frac{{Mm}}{{2r}}(2  +  3{\bar s^{00}} +{\bar s^{jk}}{\hat r^j}_{}{\hat r^k}_{}  \nonumber \\ && -3{\bar s^{0j}}v_{}^j - {\bar s^{0j}}{\hat r^j}_{}v_{}^k{\hat r^k}_{}), \label{Eq:3}
\end{eqnarray}
where the indices $j, k$ denote space coordinates, $\vec v$ the relative velocity, and $\hat r = \vec r/r$. The reference frame chosen here is the lab frame ($t$, $x^{j}=\hat{x}, \hat{y}, \hat{z}$). It has the $x$ axis point south, the $y$ axis east, and $z$ axis vertically upwards. Based on the Lagrangian, we can obtain the vertical component of the acceleration at any location on the Earth:
\begin{eqnarray}\label{equation4:eps}
g_z &=& g_0 \left(1 + \frac{3}{2}{i_1}{\bar{s}^{TT}} + \frac{1}{2}{i_4}{\bar{s}^{\hat z\hat z}}\right)- {\omega ^2}{R_ \oplus }{\sin ^2}\chi \nonumber \\
&&- {g_0}{i_4}{\bar{s}^{T\hat z}}V_ \oplus ^{\hat z} - 3{g_0}{i_1}{\bar{s}^{TJ}}V_ \oplus ^J,\label{Eq:4}
\end{eqnarray}
where $g_{0}$, ${R_\oplus }$, $V_\oplus $ are the gravitational acceleration in the absence of Lorentz violation, the radius of the Earth, and the orbital velocity of the Earth respectively. The quantities ${i_1} = 1 + i_ \oplus/3$ and $i_4 = 1 - 3i_ \oplus$ are given by $i_ \oplus = I_ \oplus/(M_ \oplus R^2_ \oplus)$, where $I_ \oplus$ is the spherical moment of inertia of the Earth and $M_\oplus$ is the mass of the Earth.

\begin{table*}[!t]
\caption{\label{tab:1} Amplitude and phase of Lorentz violation signals in the vertical gravitational acceleration in each Fourier component and the decomposed results from superconducting gravimeters data of Medicina, Italy (-5.7152 ). $\chi $ is colatitude, and $\eta $ is the inclination between the Earth's equatorial plane and orbital plane.}
\newcommand{\tabincell}[2]{\begin{tabular}{@{}#1@{}}#2\end{tabular}}
\begin{ruledtabular}
{\renewcommand{\arraystretch}{1.25}
\begin{tabular}{lccc}
\tabincell{l}{Fourier Component} &\tabincell{c}{Amplitude}&\tabincell{c}{Phase}&Measurement/${10^{ - 11}}$($1\sigma $)\\
\hline
$C_{2\omega }$ & $\frac{1}{4}{i_4}({\bar s}^{XX} - {\bar s}^{YY}){\sin }^2\chi$ & ${2\phi }$&${2.6\pm0.2}$\\
$D_{2\omega }$ & $\frac{1}{2}{i_4}{\bar s}^{XY}{\sin }^2\chi $ &${2\phi }$&${6.1\pm0.2}$\\
${C_{\omega }}$ &  $\frac{1}{2}{i_4}{\bar s}^{XZ}\sin 2\chi$      &${\phi }$&${-0.6\pm0.2}$\\
${D_{\omega }}$ &  $\frac{1}{2}{i_4}{\bar s}^{YZ}\sin 2\chi $     &${\phi }$&${-1.1\pm0.2}$\\
\hline
${{C_{2\omega  + \Omega }}}$ & $ - \frac{1}{4}{i_4}{V_ \oplus }{\bar s}^{TY}(\cos \eta  - 1){\sin }^2\chi $ & ${2\phi }$&${0.1\pm0.2}$\\
${{D_{2\omega  + \Omega }}}$ & $\frac{1}{4}{i_4}{V_ \oplus }{\bar s}^{TX}(\cos \eta  - 1){\sin }^2\chi $  & ${2\phi }$&${-0.3\pm0.2}$\\
${{C_{2\omega  - \Omega }}}$ & $ - \frac{1}{4}{i_4}{V_ \oplus }{\bar s}^{TY}(\cos \eta  + 1){\sin }^2\chi $ & ${2\phi }$&${-0.2\pm0.2}$\\
${{D_{2\omega  - \Omega }}}$ &$\frac{1}{4}{i_4}{V_ \oplus }{\bar s}^{TX}(\cos \eta  + 1){\sin }^2\chi $ & ${2\phi }$&${0.5\pm0.2}$\\
${{C_{\omega  + \Omega }}}$ & $\frac{1}{4}{i_4}{V_ \oplus }{\bar s}^{TX}\sin \eta \sin 2\chi $  &${\phi }$&${-204.8\pm0.2}$\\
${{D_{\omega+\Omega }}}$ &$-\frac{1}{4}{i_4}{V_ \oplus }[{\bar s}^{TZ}(1 - \cos \eta )-{\bar s}^{TY}\sin \eta]\sin 2\chi$ & ${\phi }$&${-112.2\pm0.2}$\\
${{C_{\omega  - \Omega }}}$ &$\frac{1}{4}{i_4}{V_ \oplus }{\bar s}^{TX}\sin \eta \sin 2\chi$ &${\phi }$&${64.5\pm0.2}$\\
${{D_{\omega  - \Omega }}}$ &$\frac{1}{4}{i_4}{V_ \oplus } [{\bar s}^{TZ}(1 + \cos \eta )+{\bar s}^{TY} \sin \eta ]\sin 2\chi$  & ${\phi }$&${37.9\pm0.2}$\\
\end{tabular}
}
\end{ruledtabular}
\end{table*}

We use the conventional sun-centered celestial equatorial reference frame ($T, X^J=\hat{X}, \hat{Y}, \hat{Z}$) \cite{16} to express the coefficients for Lorentz violation. The difference between celestial and the lab time $T$ and $t$ can be written as a phase difference $\phi  \backsimeq \omega_\oplus(t - T)$ \cite{7d}, where $\omega_\oplus$ is the angular frequency of the Earth's rotation. After appropriate transformation into the Sun-centered frame, the variation of g in Eq. (\ref{Eq:4}) can be decomposed into its Fourier components

\begin{eqnarray}\label{equation5:eps}
\frac{\delta g}{g_0} = \sum\limits_m {{C_m}} \cos ({\omega _m}t + {\phi _m}) + {D_m}(\sin {\omega _m}t + {\phi _m}),\label{Eq:5}
\end{eqnarray}
where the coefficients ${C_m},{D_m}$ are functions of coefficients ${\bar s^{\mu \nu }}$ corresponding to six frequencies $\omega_\oplus, 2\omega_\oplus, \omega_\oplus  \pm \Omega,$ and $2\omega_\oplus  \pm \Omega$. The angular frequency of the Earth's rotation and orbit are given by $\omega_\oplus  \simeq 2\pi /(23.93$h) and $\Omega  = 2\pi /(1$y). The functional forms of the coefficients $C_{m},D_{m}$ are listed in Tab. \ref{tab:1}. 

Established by Chan and Paik \cite{17}, the superconducting gravimeter presents one of the most sensitive tools to measure gravity. It converts the acceleration of a levitated test mass into a current signal in a coil. As mentioned earlier we have complied and analyzed gravity data available from IGETS \cite{18}. In order to extract Lorentz violation signals from the original superconducting gravimeters data, we need to properly separate signals caused by earthquakes and tides. The earthquake signal is at much higher frequencies than tides or Lorentz violation signal, so it can be easily identified. Depending on the length of seismic signals, they were properly handled or removed by geophysics experts or the International Center of Earth Tide (ICET) in special software like Tsoft \cite{20a}. The main focus of this work is to remove the tidal signal and look for Lorentz violation in the residual.

The observed gravity $g\left( t \right)$ could be decomposed into tides ${\rm GT}\left( t \right)$, an air pressure related term $ bP\left( t \right)$, and the Lorentz violation term $\delta {g_{LV}}$, as:
\begin{eqnarray}\label{equation6:eps}
g_{\rm meas}\left( t \right) = {\rm GT}\left( t \right) + bP\left( t \right)+\delta {g_{LV}},\label{Eq:6}
\end{eqnarray}
where $P\left( t \right)$ is atmospheric pressure and $b$ a regression parameter to be determined. ${\rm GT}\left( t \right)$ consists solid Earth tides and ocean tides, which can be expressed as $n_w$ wave groups of harmonic series:
\begin{eqnarray}\label{equation8:eps}
{\rm GT} ( t ) = \sum\limits_{n = 1}^{{n_w}} {{W_n}\!\!\sum\limits_{k = {n_s}}^{{n_e}} {{H_k}\cos \left( {{\omega _k}t + {\varphi _k}\!\! + \!\!\Delta {\varphi _n}} \right)} },\nonumber \\ \label{Eq:8}
\end{eqnarray}
where $\omega_k, \varphi_k$, and $H_k$ are the frequency, phase and amplitude of each harmonic component, respectively, given as a $\emph{priori}$ based on an Earth model. $W_n$ and $\Delta \varphi_n$ are corrections to the amplitude and phase of each tidal wave group. It is to be noted that the tides here are modeled based on conventional simple isotropic Newtonian gravity theory instead of consistent anisotropic gravity theory in SME, because the effect of anisotropy of gravity is highly negligible in tidal models based on current limits set in other works.\cite{4a, 4b, 4c, 6, 6a}.

In the empirical tidal model, the wave groups parameters ${W_n}$ and $\Delta {\varphi _n}$ are determined by fitting to observed data. A detailed treatment can be found in the ETERNA3.40 software for example (Wenzel 1996 \cite{19}), which will provide a tidal model with residual directly calculated after processing. However, as pointed out earlier, the fitted tidal model might hide potential Lorentz violation signals from being detected. But in theoretical models such as the one given by Tsoft \cite{19a}, though not as precise, the wave groups parameters are obtained by calculation independent of gravity measurement. Therefore, we use Tsoft to extract Lorentz violation signals from experimental measurements in this work.

To estimate the coefficients of the violation ${\bar s^{\mu \nu }}$ from the superconducting gravimeters data, we decompose the residual signal after subtracting the theoretical tidal model $\delta g$ into its Fourier components $(C_m, D_m)$ by least-squares fitting.  Based on the functional relations listed in Tab. \ref{tab:1}, we solve for the coefficients $\bar s^{\mu \nu}$ from these Fourier components. We present the data of Medicina, Italy as an example shown in Figure 1. The raw data spectrum after Fourier transform is plotted in Figure 1a) in the scale of cycle per day (cpd). In the same scale, we have also plotted the spectrum of theoretical tidal model in Figure 1b) with some major tidal waves labeled, the residual spectrum after subtracting the theoretical tidal model (theoretical residual) in Figure 1c), as well as the residual spectrum after subtracting the empirical tidal model (fitting model residual) in Figure 1d). All data are plotted in the unit of $\SI{1}{\micro Gal}$=$\SI{1e-8}{m/s^2}$. The Fourier amplitudes from least-squares fitting of the data at this station are listed in the last column of the Tab. \ref{tab:1}.

\begin{figure}
\includegraphics[width=0.52\textwidth]{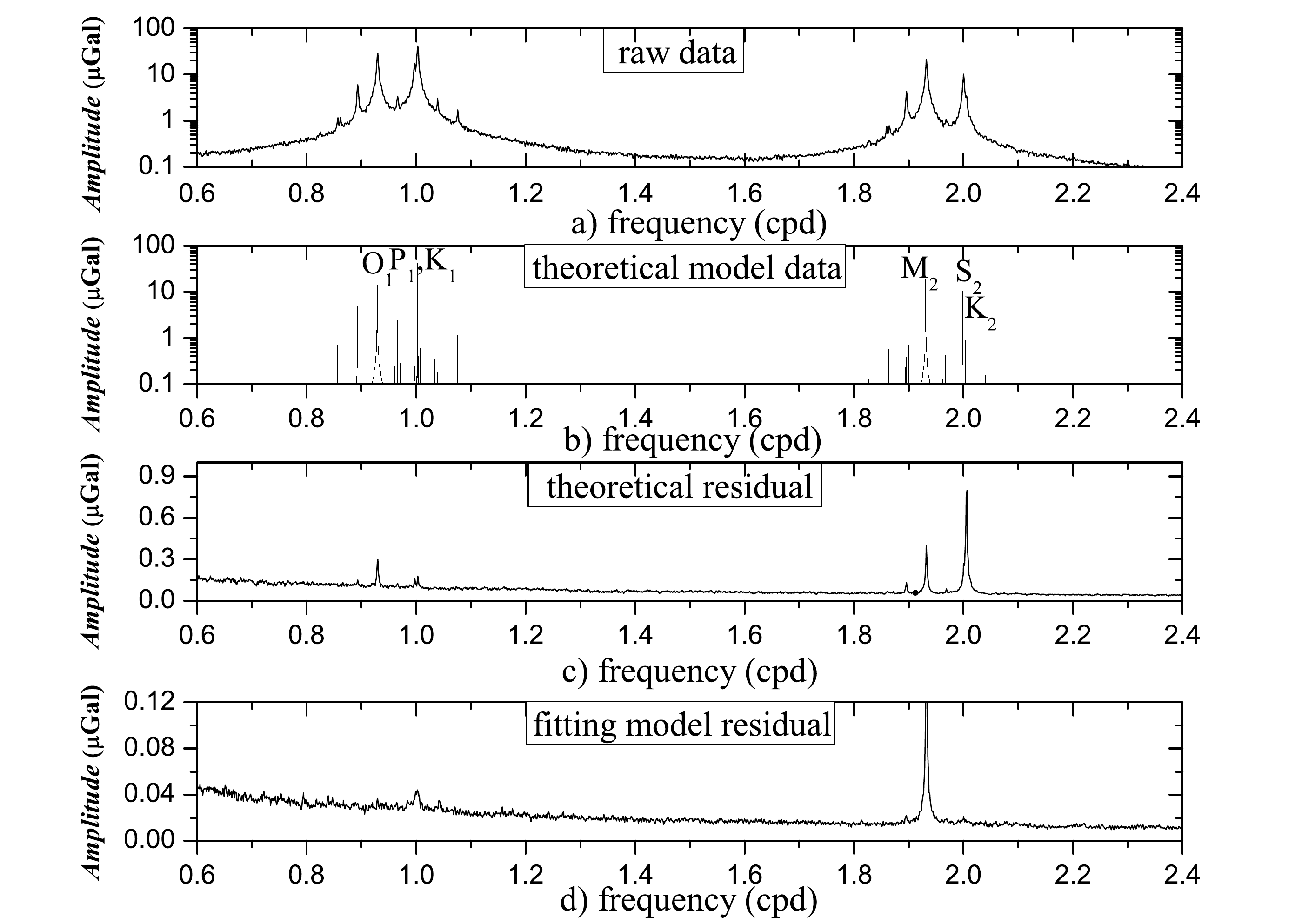}
\caption{\label{fig:1} The data of Medicina, Italy (-5.7152). This data gives an upper limit of $10^{-10}$ to the space-space components of ${\bar s^{\mu \nu }}$. Figure $a)$ shows the amplitude spectrum of raw data, figure $b)$ shows the amplitude spectrum of theoretical tidal model data, figure $c)$ shows the residual data based on the pure theoretical tidal model, and the figure $d)$ shows the residual data base on an empirical tidal model.}
\end{figure}

It is worth noting that the Fourier amplitudes at frequency $\omega\pm\Omega$ are significantly larger than others. This can be explained by the way how the theoretical tidal model works. In the theoretical tidal models like the one given by Tsoft, the tidal waves consist of main diurnal waves (${Q_1}$, ${O_1}$, ${P_1}$ ${K_1}$ etc.), main semidiurnal waves (${N_2}$, ${M_2}$, ${S_2}$ and ${K_2}$ etc.) at the leading order and other branch wave groups ($PSI_1$ etc.) \cite{20a,20b}. It predicts main tidal waves ($K_1$, $K_2$) more accurately than other tidal waves due to lack of knowledge of the Earth structure and composition at higher orders \cite{21}. In the plot in Figure 1c), the remaining spectral feature of the theoretical residual shows this inaccuracy in other waves close to the main waves. However, branch wave groups are significantly involved to predict Fourier components of tides at frequency $\omega\pm\Omega$ and $2\omega\pm\Omega$. Therefore, the tides could not be accurately subtracted at these frequencies to get time-space components of Lorentz violation signals $\bar{s}^{TX}, \bar{s}^{TY}, \bar{s}^{TZ}$. Besides, these time-space components could be constrained more accurately in astrophysics experiments like pulsar observations \cite{6, 6a}. So here we only focus on analyzing the space-space Lorentz violation components with this theoretical tidal model.

\begin{table*}[!t]
\caption{\label{tab:2} The space-space components of Lorentz violation from 12 superconducting gravimeter stations. }
\newcommand{\tabincell}[2]{\begin{tabular}{@{}#1@{}}#2\end{tabular}}
\begin{ruledtabular}
\begin{tabular}{lcccccc}
Location & $\phi$ & time & $\bar s^{XX} - \bar s^{YY}$ & $\bar s^{XY}$ & $\bar s^{XZ}$ & $\bar s^{YZ}$  \\ & rad & yr & $10^{-10}$ & $10^{-10}$& $10^{-10}$& $10^{-10}$ \\ \hline
Metsahovi, Finland &-5.4926 & 20 & ${-8.5\pm2.3}$&${-10.2\pm1.1}$&${-1.0\pm0.3}$&${-1.7\pm0.3}$\\
Medicina, Italy &-5.7152 & 19 & ${-6.9\pm1.4}$&${-9.8\pm0.7}$&${-0.1\pm0.3}$&${-2.5\pm0.4}$\\
Membach, Belgium &-5.8136& 16 & ${-18.5\pm3.6}$&${-10.6\pm1.8}$&${1.4\pm0.8}$&${-7.9\pm0.8}$\\
Sutherland, South Africa &-5.5552& 15 & ${-2.1\pm1.1}$&${-9.8\pm0.6}$&${10.5\pm0.5}$&${-3.8\pm0.5}$\\
Boulder, Co, USA &-1.4719 & 9 & ${-4.5\pm4.8}$&${-11.0\pm1.2}$&${0.6\pm0.5}$&${-2.3\pm0.5}$\\
Canberra, Australia & -3.3177 & 8 & ${-11.3\pm1.1}$&${-12.1\pm0.6}$&${8.5\pm0.5}$&${-0.3\pm0.5}$\\
Pecny, Czech Republic &-5.6603& 7 & ${-23.4\pm2.3}$&${-14.3\pm7.3}$&${-3.7\pm0.5}$&${-3.7\pm0.5}$\\
Conrad Observat., Austria &5.6416 & 7 &${-7.9\pm 2.6}$&${-9.4\pm 1.3}$&${-3.6 \pm 0.6}$&${ -0.3 \pm 0.6}$\\
Bad Homburg, Germany &-5.7681 & 7 &${ 5.8\pm 2.7}$&${-10.8\pm 1.4}$&${-3.6 \pm 0.6}$&${3.0\pm 0.6}$\\
Wetzell, Germany &-5.6936 & 6 &${-21.7\pm 3.4}$&${-9.7\pm 1.7}$&${-4.5\pm 0.8}$&${ -1.5\pm 0.8}$\\
Schiltach, Germany &-5.7731 & 4 &${ -18.5 \pm 3.6}$&${-10.6 \pm 1.8}$&${-1.4 \pm 0.8}$&${ -7.9\pm 0.8}$\\
Brasimone, Italy &-5.7244& 2 & ${-7.2\pm4.1}$&${-14.5\pm2.1}$&${-6.8\pm1.1}$&${20.7\pm1.1}$\\
\end{tabular}
\end{ruledtabular}
\end{table*}

\begin{table*}[!t]
\caption{\label{tab:3} The comparison of Lorentz violation (LV) bounds from atom interferometry \cite{4a}, the superconducting gravimeter at Bad Homburg \cite{4c} and a worldwide array of superconducting gravimeters in this work. This is the first bound of Lorentz violation obtained by a first-principles tidal model with ocean tides. The possible systematic error given here is based on residual spectrum at 2 $\omega$.}
\newcommand{\tabincell}[2]{\begin{tabular}{@{}#1@{}}#2\end{tabular}}
\begin{ruledtabular}
\begin{tabular}{lcccccc}
Coefficient&\tabincell{c}{Atom interferometry \cite{4a}}&\tabincell{c}{Superconducting\\ gravimeter at \\ Bad Homburg \cite{4c}}&\tabincell{c} {LV estimate with\\statistical errors}&\tabincell{c}{LV systematic errors\\from the tidal model}&\tabincell{c}{Overall estimate of\\LV in this work}\\
${{{\bar s}^{XX}} - {{\bar s}^{YY}}}$&${(4.4 \pm 11) \times {{10}^{ - 9}}}$&${( 2 \pm 1) \times {{10}^{ - 10}}}$&${( -8.8\pm 0.5) \times {{10}^{ - 10}}}$&${2.4 \times {{10}^{ - 9}}}$&${(-0.9 \pm 2.4) \times {{10}^{ - 9}}}$\\
${{{\bar s}^{XY}}}$&${(0.2 \pm 3.9) \times {{10}^{ - 9}}}$&${( -4 \pm 1) \times {{10}^{ - 10}}}$&${(-11.0\pm 0.3) \times {{10}^{ - 10}}}$&${1.2 \times {{10}^{ -9}}}$&${(-1.1 \pm 1.2) \times {{10}^{ - 9}}}$\\
${{{\bar s}^{XZ}}}$&${( - 2.6 \pm 4.4) \times {{10}^{ - 9}}}$&${( 0 \pm 1) \times {{10}^{ - 10}}}$&${( -3.0\pm 1.4) \times {{10}^{ - 11}}}$&$1.8 \times 10^{-10}$&${(-0.3 \pm 1.8) \times {{10}^{ - 10}}}$\\
${{{\bar s}^{YZ}}}$&${( - 0.3 \pm 4.5) \times {{10}^{ - 9}}}$&${( 3 \pm 1) \times {{10}^{ - 10}}}$&${( -2.4\pm 1.4) \times {{10}^{ - 11}}}$&$1.8 \times 10^{-10}$&${(-0.2 \pm 1.8) \times {{10}^{ - 10}}}$\\
\end{tabular}
\end{ruledtabular}
\end{table*}


We have analyzed gravity data from 12 superconducting gravimeter stations all around the world with geographic location, ocean load correction, atmospheric pressure taken into account in constructing the theoretical tidal model. There are gaps ranging from 2 days to about a year in the middle of data sets. But each continuous portion is at least a month long so the gap has no significant effect on Fourier components at frequencies  $\omega$ and $2\omega$. Following the procedure described above, we have obtained space-space components of Lorentz violation signal shown in Tab. \ref{tab:2}. Some results in this table are not compatible with one another because we have not yet included the systematic error of the tidal model into this analysis. In Tab. \ref{tab:3}, we combined all data to get an overall estimation of Lorentz violation by least-squares fit. The mean values and statistical errors of Lorentz violation space-space components are listed in the 4th column of Tab. \ref{tab:3} along with results from previous work. In order to estimate systematic error, we analyzed the residual spectrum after subtracting the theoretical model at frequencies of interest ($\omega$ and $2\omega$), as shown in the Figure 1c) as an example. Since the theoretical model always predicts the main tidal waves (K1, K2) at $\omega$ and $2\omega$ better than other close-by waves as shown in the spectral feature of theoretical residual in the same plot, we can use the largest residual peak at Lorentz-violation-free frequencies close by to set conservative systematic bounds for the theoretical model at each main tidal wave respectively.  Based on least-squares fitting of residual data for all 12 stations, we assigned $\SI{0.15}{\micro Gal}$ and $\SI{0.5}{\micro Gal}$ as conservative estimate for potential tide model error at frequencies $\omega$ and $2\omega$ respectively. The corresponding systematic error for each Lorentz violation component is listed in the fifth column of Tab. \ref{tab:3}. Thus we report the overall estimate of space-space Lorentz violation components in the last column of Tab. \ref{tab:3}. As we see, our limits for the space-space components of Lorentz violation are up to the level of $10^{-10}$, one order of magnitude smaller than that of atom interferometry \cite{4a}, and similar to the single superconducting gravimeter result listed in \cite{4c}. And more than precision, we have also resolved the accuracy issue caused by inaccurate tidal model used at the precision of superconducting gravimeters.

In this letter, we have tested the local Lorentz invariance and resolved an issue of accuracy by analyzing gravity measurement of a worldwide array of superconducting gravimeters with a first-principles tidal model with ocean tides included for the first time and reached a competitive result of ground-based experiments to bound space-space Lorentz violation components to up to  $10^{-10}$. The ground-based gravimetry experiments for detecting Lorentz violation are primarily limited by inadequate accuracy in all available theoretical tidal models. In the future, the sensitivity to Lorentz violation coefficients would potentially be improved by orders of magnitude with improved tidal models. Moreover, the approach in this work of using a worldwide network of superconducting gravimeters and correctly handling tides with first-principles tidal models, will also open new ways of testing exciting new gravitational physics like some dark-matter theories beyond the Standard Model.

We gratefully acknowledge support by the National Natural Science Foundation of China (Grants No.91636221 and No.91636219). Besides, this material is also based upon work supported by the National Science Foundation under CAREER Grant No. PHY-1056620 and the David and Lucile Packard Foundation, National Aeronautics and Space Administration Grants No. 041060-002, 041542, 039088, 038706, and 036803.

\end{document}